\documentclass[twocolumn, graphics, floatfix, a4paper, aps, prl, longbibliography, superscriptaddress]{revtex4-2} 
\usepackage{graphicx}
\usepackage{bm}
\usepackage[usenames,dvipsnames]{xcolor}
\usepackage[colorlinks=true,urlcolor=blue,citecolor=blue,linkcolor=blue]{hyperref}
\usepackage{amssymb,amsmath}
\usepackage[normalem]{ulem}
\usepackage{booktabs}
\usepackage{longtable}
\usepackage{braket}
\usepackage[T1]{fontenc}
\usepackage[utf8]{inputenc}
\usepackage[english]{babel}
\usepackage{xspace}
\usepackage{epstopdf}
\usepackage[final]{microtype}
\usepackage{enumitem}
\usepackage{subfigure}
\usepackage{amsmath}

\graphicspath{{./Figures/}}

\begin{document}
\title{Essay: Where Can Quantum Geometry Lead Us?}

\author{P\"aivi T\"orm\"a}
\email{paivi.torma@aalto.fi}
\affiliation{Department of Applied Physics, Aalto University School of Science, FI-00076 Aalto, Finland}

\begin{abstract}
Quantum geometry defines the phase and amplitude distances between quantum states. The phase distance is characterized by the Berry curvature and thus relates to topological phenomena. The significance of the full quantum geometry, including  the amplitude distance characterized by the quantum metric, has started to receive attention in the last few years. Various quantum transport and interaction phenomena have been found to be critically influenced by quantum geometry. For example, quantum geometry allows counterintuitive flow of supercurrent in a flat band where single electrons are immobile. In this Essay, I will discuss my view of the important open problems and future applications of this research topic and will try to inspire the reader to come up with further ideas. At its best, quantum geometry can open a new chapter in band theory and lead to breakthroughs as transformative as room-temperature superconductivity. However, first, more experiments directly showing the effect of quantum geometry are needed. We also have to integrate quantum geometry analysis in our most advanced numerical methods. Further, the ramifications of quantum geometry should be studied in a wider range, including electric and electromagnetic responses and interaction phenomena in free- and correlated-electron materials, bosonic systems, optics, and other fields.  

\textit{Part of a series of Essays in Physical Review Letters which concisely present author visions for the future of their field.}

\end{abstract}

\maketitle

\textit{The concept of quantum geometry.}---In quantum physics, eigenvalues and eigenstates fully describe the physical behavior of a system. For a long time, emphasis was placed on the eigenvalues since they give the observable quantities: energies, momenta, spin, and so on. The eigenfunctions give the probabilities of finding the system in a certain configuration (position, momentum, etc.), often of only indirect importance, e.g., via calculation of expectation values and transition rates. Of course, this has dramatically changed in recent times. A notable example is entanglement, which is an inherently wave function or eigenstate property and now forms the fundamental resource of quantum information science and technology. Another one is topological physics, which deals with structural properties of the eigenfunctions. Now, it seems that topological physics was perhaps only one aspect of a wider and possibly even more influential concept, namely, quantum geometric physics. 

Quantum geometry defines the geometry of the eigenstate space~\cite{Resta2011}. As in the classical world, the geometry of a space determines distances, for example the distance between two points is different on a plane and on a sphere. Likewise, the distances between quantum states depend on the geometry of the eigenstate space, and this is captured by the quantum geometric tensor (QGT)~\cite{Provost1980} (or Fubini-Study metric) ${\cal{B}}_{ij}(\mathbf{k})$:
\begin{equation}\label{def:qgt}
		{\cal{B}}_{ij}(\mathbf{k})=\langle\partial_{i}u_{\mathbf{k}}|\partial_{j}u_{\mathbf{k}}\rangle-\langle\partial_{i}u_{\mathbf{k}}|u_{\mathbf{k}}\rangle\langle u_{\mathbf{k}}|\partial_{j}u_{\mathbf{k}}\rangle,
	\end{equation}
	where $\partial_{i}\equiv\partial/(\partial k_{i})$ with $i=x,y$ and $u_{\mathbf{k}}$ is a wave function parametrized by a quantity $\mathbf{k}$ (which could be, for example, the lattice momentum).
Its real part, the quantum metric $\Re{\cal{B}}_{ij}(\mathbf{k})\equiv g_{ij}$ tells about the orthogonality, i.e., amplitude distance of quantum states under small changes. The last term of Eq.~(\ref{def:qgt}) is real due to normalization, thus the imaginary part of the QGT is $\Im{\cal{B}}_{ij}(\mathbf{k}) = -i(\langle\partial_{i}u_{\mathbf{k}}|\partial_{j}u_{\mathbf{k}}\rangle - \langle\partial_{j}u_{\mathbf{k}}|\partial_{i}u_{\mathbf{k}}\rangle)/2$. From this one can see that $\Im{\cal{B}}_{ij}(\mathbf{k})$ is the well-known Berry curvature (defined in vector form as $i \nabla_\mathbf{k}\times \langle u_{\mathbf{k}}|\nabla_\mathbf{k} u_{\mathbf{k}}\rangle$), which provides information about changes of the eigenstate phase (for more information see the papers~\cite{Torma2022,Peotta2023review}). As the integral of the Berry curvature gives the Chern number, the QGT contains information about the system topology too. These concepts have been long known~\cite{Provost1980,Souza2000,Resta2011}. The idea that they can critically affect physical properties in interacting many-body systems is a more recent development~\cite{Parameswaran2013,Neupert2015,Liu2023,Torma2022}. 

Most quantum geometry studies in condensed matter physics have focused on the geometry of the eigenstates of a Bloch energy band in a periodic lattice system (solid state, optical, or other~\cite{Parameswaran2013,Neupert2015,Liu2023,Rossi2021,Torma2022,Peotta2015}). Historically, the structure of the Bloch energy bands has enabled in a simple way the classification of matter into insulators, metals, semiconductors, and semimetals~\cite{scalapino1993}. Quantum geometry, on the other hand, gives information  about the structure of the Bloch functions in a band. It only becomes nontrivial if the band consists of contributions of different orbitals (see Fig. 1), i.e., when the multiband (multiorbital) nature of the problem is important ("orbital" refers here to a general degree of freedom, e.g., atomic orbital, spin, or light polarization). Since the Wannier functions of a band are given by the Bloch functions via a Fourier transformation, it is not difficult to imagine that quantum geometry of the band actually has a relation to the Wannier functions too. Moreover, this is an important one: quantum geometry determines the localization properties of the Wannier functions. Overlaps of the Wannier functions of nearby lattice sites affect nearly all transport and interaction phenomena, so it is no wonder that quantum geometry is emerging as a fundamental and powerful concept for understanding solid state and other periodic systems. 

\begin{figure}
    \centering
    \includegraphics[width = \columnwidth]{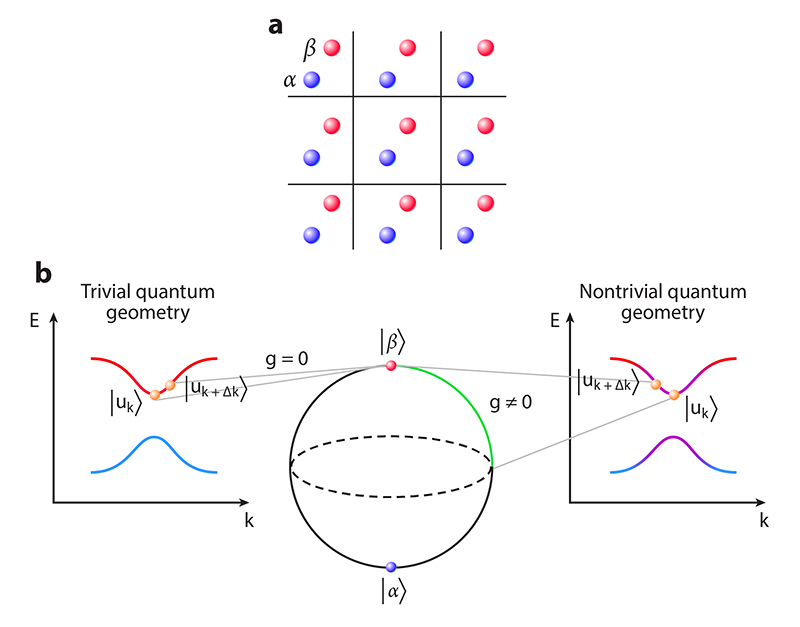}
    \caption{(a) Atoms in materials may form a lattice that the electrons "feel." Each unit cell of the lattice can have several orbitals, labeled $\alpha$ and $\beta$ in our example, which only has two. The orbital states $|\alpha \rangle$ and $|\beta \rangle$ are orthogonal and can be illustrated by the Bloch sphere, (b), middle. The states of the electrons in a band may involve one orbital only, indicated by a single color in (b), left. The distance between two states when the change in the lattice momentum $\delta \mathbf{k}$ is infinitesimally small  then vanishes because the orbital state remains the same. The quantum geometry is trivial in this case and the quantum metric $g=0$. In the nontrivial case (b), right, the electron state in a band can be a superposition of two orbitals ("mixed" color). Then, upon a change in $\mathbf{k}$ that changes this superposition, the new state will have a finite quantum distance from the original one due to the orthogonality of the orbital states; the quantum metric is nonzero. For a simple model example, see~\cite{Cuerda2023}.}
    \label{fig:figure1}
\end{figure}

\textit{Relevance of quantum geometry to physical phenomena.}---It has been theoretically predicted that quantum geometry governs a variety of physical phenomena. In the case of free (or more precisely, noncorrelated) electrons in band insulators and semimetals, quantum geometry appears as an important ingredient of the Hall effect, shift currents, circular photogalvanic effect, and resonant optical responses, just to name a few examples. Review articles on these topics do no exist (a pressing to-do task for the experts in the field), and I choose not to attempt to give credit to the large amount of original and seminal theory work in this and other areas mentioned in this Essay (a few references can be found in~\cite{Huhtinen2023,Torma2022,Huhtinen2022,Ahn2022}). Concerning (strongly) correlated electron systems, there are, so far, fewer examples of phenomena essentially governed by quantum geometry: the most prominent ones are fractional Chern insulators~\cite{Parameswaran2013,Neupert2015,Liu2023} and flat-band superconductivity~\cite{Rossi2021,Torma2022,Peotta2023review}. In the latter case, quantum geometry beautifully solves an outstanding puzzle: how could there be superconductivity in a flat band, if electrons are, due to their flat dispersion, localized? It turns out that quantum geometry, which guarantees sufficient overlap of Wannier functions, enables this~\cite{Peotta2015}, with potentially remarkable consequences since now the diverging density of states of a flat band can be utilized to achieve higher critical temperatures. Here it is good to reflect on the fact that the \emph{single-particle} quantum metric gives essential information about an interacting, correlated many-body state. This boils down to the role of the Wannier functions in our theoretical description of interacting phases. 

In flat-band superconductivity as well as in many other contexts, the role of quantum geometry is nicely illustrated by the current operator of a multiband system ($m$,$n$ are band indices and $i=x,y,z$): $\bra{m}j_i\ket{n} = \delta_{mn} \partial \epsilon_n/\partial k_i + (\epsilon_m - \epsilon_n) \bra{\partial m/\partial k_i}n\rangle$, where $\mathbf{k}$ is the momentum and $\epsilon(\mathbf{k})$ gives the dispersion. The first term is the conventional current arising from the group velocity, and the latter leads to the quantum geometric effects. From this formula it is obvious that quantum geometric effects are of multiband nature (since the latter term is nonzero only for different bands, $m\neq n$) and that they dominate in a (nearly) flat band (since then the intraband terms ($m=n$) vanish as $\partial \epsilon_n/\partial k_i \sim 0)$. However, quantum geometry can be qualitatively and quantitatively significant also for bands with a considerable kinetic energy. Given this realization, what should we do to best unveil the potential of quantum geometric physics? 

\textit{Experiments: Present and future.}---First and foremost, since physics is an experimental science, there should be many more experimental demonstrations of the significance of the quantum metric part of quantum geometry to physical phenomena (for the Berry curvature and Chern number part, there is already a remarkable amount of experimental work, for example, with topological insulators).  

Naturally, we should try to make the characterization of the full QGT a routine measurement available for any given physical system. As we refer here to the quantum geometry of the noninteracting (or weakly interacting, noncorrelated) bands, such measurements are relevant only in cases where one can effectively turn the interactions or correlations off by some means, for example, by temperature, magnetic or electric field, doping, density, absence of nonlinear medium, and so forth. Various high-frequency responses as well as tomographic approaches can be used for measuring the QGT, and experimental observations of the full QGT already exist for qubits in diamonds~\cite{YuNatSciRev2019}, superconducting circuits~\cite{ZhengChinPhysLett2022}, ultracold atom~\cite{Yi2023} and polariton systems~\cite{Gianfrate2020}, as well as plasmonic lattices~\cite{Cuerda2023}. For example, the QGT related to light polarization can be obtained by a tomographic approach where six different polarizations are measured at each $\mathbf{k}$~\cite{Gianfrate2020,Cuerda2023}. However, more methods, tailored for different physical contexts, for example, various new 2D quantum materials, are needed.   

In the long run, what really matters is that the quantum geometry of the band can affect and control other physical phenomena. There are already a few intriguing directions of experimental work on this. In polariton systems, it has been shown that quantum metric affects the anomalous Hall drift~\cite{Gianfrate2020}.
The nonlinear Hall effect induced via quantum metric by interfacing even-layered MnBi$_2$Te$_4$ with black phosphorus has been observed recently~\cite{Gao2023nonlinHall}. Similar experiments on  other predicted quantum geometric transport phenomena in noncorrelated systems are likely to appear soon, considering the vast possibilities offered by layered and other 2D materials. For maximizing the chance of future applications, it is important that such studies are conducted for bulk materials as well. 

Concerning correlated electron systems,
the recent experimental advances on fractional Chern insulators in twisted graphene and MoTe$_2$~\cite{Xie2021fractional,Cai2023fractional,Park2023fractional} now open the way for one to study quantum geometry effects in these materials in detail, for example, the role of Berry curvature distribution in the Brillouin zone. 
The superfluid weight (stiffness) in flat bands has been predicted to be provided by the Brillouin-zone-integrated quantum metric~\cite{Peotta2015}; recently, the superfluid weight was estimated via the critical current and critical field measurements of twisted bilayer graphene, indicating that quantum geometry plays a major role~\cite{tian2023}.  

Starting from these promising developments, experiments in which the role of quantum geometry is precisely and unambiguously defined are needed. In the case of complex correlated systems, this requires deep theoretical analysis also. The conventional and quantum geometric contributions typically combine, so distinguishing them requires care. Fortunately, they often scale differently with system parameters such as density, interactions, and temperature, as one can anticipate, e.g., from the very different types of contributions to the current operator: $\bra{m}j_i\ket{n} = \delta_{mn} \partial \epsilon_n/\partial k_i + (\epsilon_m - \epsilon_n) \bra{\partial m/\partial k_i}n\rangle$.

Once a few smoking-gun experiments on quantum geometry effects have been achieved, the focus should swiftly shift from showing that "quantum geometry is there" to how can we utilize this concept to better understand nature and, eventually, to create new technologies.

\textit{Updating computational methods.}---A large amount of condensed matter and materials research is based on widely used computational methods, for example, density functional theory, quantum Monte Carlo technique, and dynamical mean-field theory. To proceed on the quantum geometry road, we should implement the extraction---and smart visualization----of quantum geometric concepts as a standard functionality of these tools. This is easier said than done. First, it is important to understand which quantities are the most relevant to extract from the numerically obtained data: the quantum metric and Berry curvature everywhere on the Brillouin zone, or just Brillouin-zone-integrated quantities? Or perhaps just directly some information about the Wannier functions? How should we obtain that information accurately and computationally efficiently, keeping in mind that these numerical methods inevitably contain some approximations or limitations? In (strongly) interacting systems, quantum geometric transport and interactions emerge when projecting interactions defined with a large set of bands, that is, the full system, down to a low energy band. Such downfolding or projection requires extreme care. Despite challenges, our goal should be that the most powerful numerical methods of condensed matter physics will provide the essentials of the quantum geometry of the bands, properly visualized, as easily as they give the energies. Once this becomes a routine, we will start seeing things from the perspective of quantum geometry, like we now do from the band structure's, and this will be a source of understanding and discovery. 

Perhaps algorithms will see even more. Machine learning is becoming increasingly important in materials discovery~\cite{Schmidt2019}, and incorporating quantum geometric quantities in search of new materials is worth considering. For instance, one might hint to the algorithm that not only a flat band is good for high critical temperature superconductivity but also a suitable type of quantum geometry. However,  I have heard from experts that the algorithms usually develop best “on their own". Yet, we should try finding out if quantum geometric concepts will help with machine learning for materials, because potential discoveries on that front could be thrilling. 

\textit{The annoying necessity of nitpicking.}---It is said that the devil is in the details, but I would rather say that: \textit{the devil is in the supplementary.} The quantum geometric tensor is gauge invariant; thus it is measurable. However, it is basis dependent. To clarify this, let us consider a lattice tight-binding model with multiple orbitals, somewhat like the situation shown in Fig.~1. If one alters the physical positions of the orbitals, while keeping everything else such as hoppings fixed (a bit unphysical, but in a model system one can do it), many macroscopic quantities and responses remain the same, including the superfluid weight. However, the quantum metric and the Berry curvature change! The resulting discrepancy for the connection of the superfluid weight and the quantum metric in flat band superconductivity was missed in a large body of literature, until in Ref.~\cite{Huhtinen2022} it was found out that one should use the minimal quantum metric, a basis-independent quantity defined through symmetry. The devil was indeed in the supplementary information of the original work that discovered quantum geometric superconductivity~\cite{Peotta2015}, where the superconducting order parameters were assumed real in the presence of the supercurrent, which can be safely done in suitably symmetric systems but not in general. Self-consistent evaluation of the order parameters in the presence of the current is the key; now it has been shown that the minimal quantum metric result~\cite{Huhtinen2022} can also be obtained from a random phase approximation analysis of the superfluid response~\cite{Tam2023}. The basis-dependence issue should be kept in mind in future theoretical work. Further, much of the quantum geometry in condensed matter literature, also beyond flat-band superconductivity, should be revisited in this sense. In some cases the physical observable in question may indeed depend on the basis, in others not, and then one cannot use quantum geometry in a naive way. Moreover, the intuitive understanding of the minimal quantum metric should be worked out, starting perhaps from the finding that it emerges in a natural way from the two-body problem in a flat band~\cite{Huhtinen2022}. I wonder whether it would be possible to formulate a basis-independent description of some physically relevant essential features of quantum geometry, in a similar spirit as Provost and Vallee introduced the QGT as a gauge-invariant way of measuring quantum distances~\cite{Provost1980}.   

Another dangerous pitfall is related to the fact that quantum geometry effects are the most prominent in flat bands, and thus while hunting them, one frequently enters the land of the missing Fermi surface. It is amazing how much of the condensed matter physics theory describing quantum states and their responses or excitations is done utilizing, explicitly or implicitly, the existence of the Fermi surface. It is frequently assumed that relevant phenomena happen only around the Fermi surface, low momentum states are Pauli blocked, and nasty divergences at the Fermi surface are negligible under integrals over the whole momentum space; see Ref.~\cite{Huhtinen2023} for inspiring examples. Yet, I believe it will be possible to rework most of these treatments to find predictions for the flat band case---and the physics will be excitingly different! A nice example is the flat-band version of the Cooper problem~\cite{Torma2018}, where the effect of the two-body interaction is now to give the Cooper pair a finite effective mass and mobility instead of destabilizing the Fermi sea. 

\textit{Widening quantum geometry.}---Quantum geometry is a broader concept than the single-particle (noninteracting) QGT that I have been discussing. First, the quantum distance itself ($\sqrt{1-|\langle \phi | \psi \rangle|^2}$, where $\phi$ and $\psi$ are two quantum states), instead of its infinitesimal version, the quantum metric, can be relevant as has been already shown in the case of flat-band Bose-Einstein condensate excitations. Moreover, work on other quantum geometric quantities, such as Christoffer symbols and Riemann curvature tensors, to describe physical phenomena has already started~\cite{Ahn2022} and should be continued. 

In nearly all condensed-matter related quantum geometry work, the QGT has been parametrized by the Bloch (lattice) momentum. One can, however, define also a "local quantum metric"~\cite{Torma2018} characterizing distances of wave functions in real space (by the way, this one is basis independent); interestingly, this quantity turned out to be relevant for the flat-band Cooper problem~\cite{Torma2018} and, quite surprisingly, for the quantum geometric effects of electron-phonon coupling~\cite{yu2023nontrivial}. Yet, such a local quantum metric is all but unexplored, and perhaps there are more  modified versions of the QGT (including time-dependent ones), which have great potential in  explaining and characterizing physical behavior.   

The positive definiteness of the QGT allows us to derive fundamental bounds for quantities that depend on quantum geometry, for example, the superfluid weight in a flat band is lower bounded by the Chern number~\cite{Peotta2015}. For other topological invariants see Ref.~\cite{Torma2022}. There are probably many more connections to be found between physical observables and band structure and topological invariants (known and new ones), in particular when utilizing new concepts such as the minimal quantum metric~\cite{Huhtinen2022} and going beyond the highly symmetric (with respect to time reversal, rotation, etc.) cases for which the present bounds have been derived.

One can also define the QGT for an interacting many-body state: the so-called many-body QGT~\cite{Souza2000}. To defend the single-particle QGT, I must immediately remind the reader that its ability to provide important information about a (correlated) many-body system is powerful precisely because we can calculate it without major difficulties with our currently available methods. However, we should think about the future: maybe quantum computers will become available for fully quantum simulations, at least for intermediate size systems. Then the calculation of the many-body QGT will be feasible beyond few-particle systems. Researchers who currently run problems on the existing quantum computers may explore whether and how the many-body QGT influences physical phenomena (in particular, emergent phenomena arising from strong correlations), while others can study the role of the many-body quantum metric at a general level and in systems tractable by exact diagonalization, density-matrix renormalization-group, tensor networks, and similar methods.

One very important research direction is to develop descriptions of quantum geometry effects for the case of touching bands, for example a Dirac cone touching a flat band---not an untypical scenario. Namely, the quantum metric diverges at band touchings and Berry curvature is ill defined too. Yet it has been numerically shown that band touchings enhance the critical temperature of flat-band superconductivity~\cite{Huhtinen2022,Torma2022}. A whole new theory framework is needed to capture the physics-relevant quantum geometry aspects even in cases where the standard concepts cannot be used. Inspiration for this search can be found from the effective mass tensor of a two-body bound state in a system of multiple nonisolated bands where the quantum geometric contributions and band dispersions intertwine~\cite{Iskin2022}. 

Finally, widening means also that we condensed-matter--quantum-geometry enthusiasts should make friends with the communities that have been studying quantum geometry for a long time---high energy physics, cosmology, and quantum information---and dive into their literature~\cite{Liang2023QGreview}. The danger of drowning there is of course considerable, but the embarrassment of reinventing too many wheels is an even more daunting prospect. Those communities might also learn something from us, or at least get amazed by how nature, once again, beautifully follows abstract mathematical concepts~\cite{Wigner1960}.

\textit{Exploring physical phenomena with the quantum geometry perspective.}---We should search for more contexts where quantum geometry is relevant. In the beginning, it is fine enough to just identify that quantum geometry plays a role in some physical phenomena. In the long run, we should develop an understanding and become intuitive about the typical ways quantum geometry works, and then utilize that information to design the desired systems and behavior, i.e., to take the first steps toward engineering. 

As described above, it is already well understood that nontrivial quantum geometry can facilitate transport and prominently so in (almost) flat bands where kinetic energy vanishes. Quantum geometry effects on various electronic transport and optical responses have been identified, but there is plenty of room for more work, in particular, for cases where the electrons are strongly interacting. Concerning correlated ground states of matter, one can go beyond fractional Chern insulators and flat-band superconductors where quantum geometry effects have already been predicted. One obvious area is magnetism: many of the flat-band superconductivity results can be mapped to magnetism in particle-hole symmetric systems. A big breakthrough would be to understand whether and how quantum geometry, in general, affects the competition between various interacting and correlated phases: It would be fantastic if we could conclude or even guess based on quantum geometry whether a charge density wave, magnetic phase, superconductor, or something else will win the game. Most examples so far are about phases of matter enhanced by quantum geometry, but it could also be detrimental for some phases, e.g., via enhancing fluctuations. It is intriguing and somewhat surprising that quantum geometry was recently found to significantly affect electron-phonon coupling, even in dispersive-band systems such as graphene and MgB$_2$~\cite{yu2023nontrivial}. Inspired by such results, quantum geometry effects in various microscopic interaction mechanisms is an important topic for future study and currently almost unexplored. Concerning flat-band superconductivity, the most significant future direction is to increase the critical temperature of superconductivity. In a flat band, the critical temperature is linearly proportional to the interaction, not exponentially suppressed, and quantum geometry provides the supercurrent. However, the critical temperature depends on quantum geometry, and more work is needed to understand how so that we can maximize the temperature. Thereafter, we need to find materials with suitable flat bands and quantum geometries. 

Quantum geometry is likely to influence bosonic systems quite differently from fermionic ones, at least in the weakly interacting limit. This is due to the tendency of bosons to occupy a single quantum state at low temperatures. Since quantum geometry is about distances between states, one might wonder whether it matters at all if only one state is populated. Quantum geometry, however, does control bosonic \emph{excitations} in an \emph{interacting} system and dominantly so if the band is flat. There is a vast amount of work to be done on the role of quantum geometry in bosonic systems, even in the weakly interacting case, since flat-band systems with suitable quantum geometry offer unique opportunities to study beyond mean-field physics (for examples, see~\cite{Torma2022,Peotta2023review}). And the strongly interacting limit is almost untouched. It is urgent that we find more contexts in which bosonic quantum geometry effects can be studied through experiments, to get a firm basis for this emerging topic.

The QGT, despite the word \emph{quantum} in its name, also describes the distances between solutions of classical wave equations. Therefore, quantum geometry studies are relevant in the domain of classical optics, acoustics, and any fields dealing with waves; indeed work has already begun, see e.g.~\cite{Cuerda2023} and references therein. (To be precise, I should replace \emph{quantum geometry} by \emph{eigenmode geometry}, but nitpicking is not a necessity here.) With light, the "orbital" degree of freedom can be the polarization. Thus quantum geometry can be used for understanding and designing polarization properties of light. On the other hand, photonic lattice structures with multiple orbitals in the unit cell can be fabricated, providing another route for quantum geometry studies. I expect the most interesting and useful results to come from the combination of optical nonlinearities and quantum geometry. 

I also believe that classical, in particular optical, systems are the best ones to enter the world of \emph{non-Hermitian} quantum (or eigenmode) geometry. The QGT for the non-Hermitian case has been defined, but there are various possibilities for other definitions based on different combinations of the left and right eigenstates~\cite{Brody20132,Cuerda2023}, and experiments are needed to guide the way. Experimental studies of non-Hermitian interacting \emph{quantum} systems may be challenging while the theory is not fully clear. Therefore, classical optical systems that can be microscopically simulated with great accuracy and studied experimentally with high precision may provide the best early progress for non-Hermitian quantum geometric physics.     

\textit{Transformative impact of quantum geometric physics.}---Last but not least, we should use this new concept to find something truly significant for humankind. Most probably Felix Bloch did not understand what would follow from pointing out that the lattice momentum is a good quantum number and that one could use certain eigenfunctions, later named Bloch functions. He likely had no idea, based on these concepts, that one day the band theory of solids would be formulated and from that would stem the understanding that led to the wide usage of semiconductors. Bardeen, Brattain, Shockley, and their Bell Labs co-workers would not have been able to–-at least in my humble view-–-develop the transistor without the insight and conceptual tool that band theory gives and without the work that had been done with band theory before them. And without the transistor, our world would not be the same. Band theory made a difference. It remains to be seen whether quantum geometry, including both quantum metric and the Berry curvature and thus topological physics, will play an equally large role in our world. What is clear is that the world needs out-of-the-box discoveries: world energy production and consumption are literally burning questions, as are scarcity of materials and several other gut-wrenching problems. Innovations as big as the transistor are much more urgently needed now than at the time transistor was invented. Therefore, we should try all possibilities. Perhaps the most notable prospect from quantum geometry is to guide the way to room temperature superconductivity. The interplay of light, electronic transport, and quantum geometry intrigues me as well, because photovoltaics is another area where major technological advances can have world-changing impact. There may be other equally important goals for which quantum geometry can provide guidance, but my imagination stops here.  I give the floor to the reader who probably has a different background and can thus come up with different ideas. 

\begin{acknowledgments}
P. T. would like to acknowledge discussions with Milan Allan, Andrei Bernevig, Dmitri Efetov, Ion Errea, Pertti Hakonen, Kristjan Haule, Tero Heikkil\"a, Kukka-Emilia Huhtinen, Miguel Marques, Andrew Millis, and Sebastiano Peotta, and funding from the Academy of Finland under Project No.~349313, as well as funding from the SuperC Collaboration through the Jane \& Aatos Erkko Foundation, Keele Foundation, and Magnus Ehrnrooth Foundation.

\end{acknowledgments}

\begin{figure}
    \centering
    \includegraphics[width = 0.7\columnwidth]{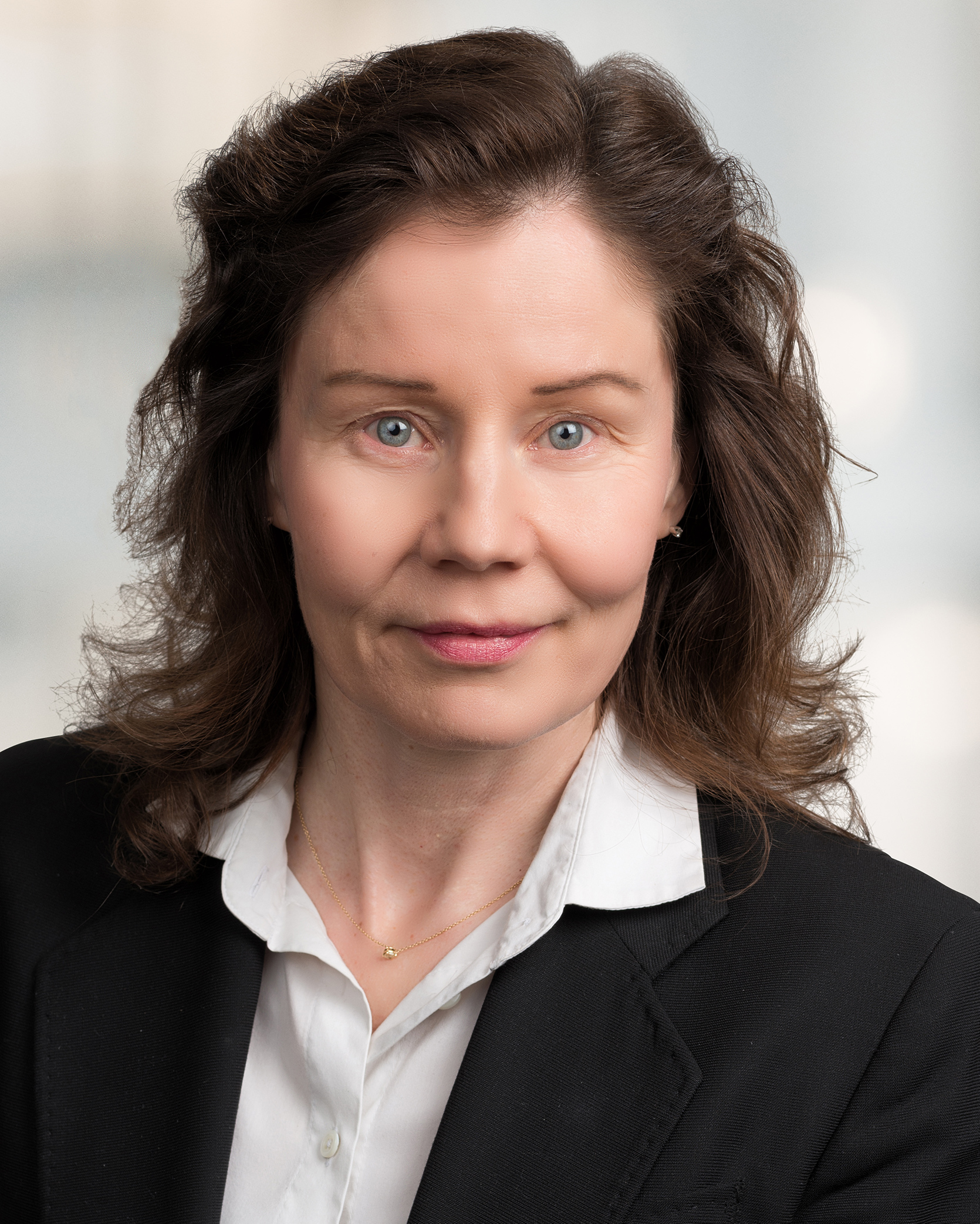}
    \caption{Päivi Törmä is a professor of physics at Aalto University, Finland. She has a MSc degree from the University of Oulu, Finland, a Master of Advanced Study degree from the University of Cambridge, U.K., and a PhD in 1996 from the University of Helsinki. Her research ranges from theoretical quantum many-body physics to experiments in nanophotonics. Recently her work has focused on the role of quantum geometry in superconductivity, especially in flat bands. Professor Törmä has given over 150 invited talks, published over 200 peer-reviewed articles, and is an elected member of the Academia Europaea.  }
    \label{fig:figure1}
\end{figure}

\bibliography{references}
\end{document}